\documentclass[sigconf]{acmart}

\copyrightyear{2027}
\acmYear{2027}
\setcopyright{cc}
\setcctype{by-nc-nd}
\acmConference[WSDM 2027] {Proceedings of the Twentieth ACM International Conference on Web Search and Data Mining}{February 15--19, 2027}{Hong Kong, Hong Kong.}
\acmBooktitle{Proceedings of the Twentieth ACM International Conference on Web Search and Data Mining (WSDM 2027), February 15--19, 2027, Hong Kong, Hong Kong}
\acmISBN{979-8-4007-3117-4/2027/02}
\acmDOI{10.1145/3862322.3863712}

\usepackage{booktabs}
\usepackage{graphicx}
\usepackage{multirow}
\usepackage[table]{xcolor}
\usepackage{graphicx}
\definecolor{mygray}{gray}{0.9}
\usepackage{adjustbox}

\usepackage{pgfplots}
\pgfplotsset{compat=1.18}

\acmSubmissionID{pfd0366}

\begin{document}

\title{IntentGR: Enhancing Intent for End-to-End Personalized
Generative Retrieval in Search}

\author{Gefan Ye}
\correspondingauthor
\authornote{This work was fulfilled when Gefan Ye interned at Alibaba International Digital Commercial Group.}
\affiliation{%
  \institution{Zhejiang University}
  \city{Hangzhou}
  \country{China}}
\affiliation{%
  \institution{Alibaba International Digital Commercial Group}
  \city{Hangzhou}
  \country{China}}
\email{gefanyeh@zju.edu.cn}

\author{Yanjing Wu}
\affiliation{%
  \institution{Alibaba International Digital Commercial Group}
  \city{Hangzhou}
  \country{China}}
\email{qinge.wyj@alibaba-inc.com}

\author{Yinfu Feng}
\affiliation{%
  \institution{Alibaba International Digital Commercial Group}
  \city{Hangzhou}
  \country{China}}
\email{fyf200502@gmail.com}

\author{Zhiqiang Wang}
\affiliation{%
  \institution{Alibaba International Digital Commercial Group}
  \city{Hangzhou}
  \country{China}}
\email{zhoufan.wzq@alibaba-inc.com}

\author{Liang Li}
\affiliation{%
  \institution{Alibaba International Digital Commercial Group}
  \city{Hangzhou}
  \country{China}}
\email{liliang1037@gmail.com}

\author{Jun Xiao}
\affiliation{%
  \institution{Zhejiang University}
  \city{Hangzhou}
  \country{China}}
\email{junx@cs.zju.edu.cn}

\author{Rong Xiao}
\affiliation{%
  \institution{Alibaba International Digital Commercial Group}
  \city{Hangzhou}
  \country{China}}
\email{xiaorong.xr@taobao.com}

\author{Xiaoyi Zeng}
\affiliation{%
  \institution{Alibaba International Digital Commercial Group}
  \city{Hangzhou}
  \country{China}}
\email{yuanhan@taobao.com}

\renewcommand{\shortauthors}{Gefan Ye et al.}

\begin{abstract}
    Personalized generative retrieval in search should jointly satisfy
    the user's immediate search intent and long-term preference. Recent
    end-to-end approaches, primarily explored in recommendation, unify
    item tokenization and sequence generation within a single model, allowing semantic identifiers (SIDs) to be optimized toward the
    downstream generation objective. However, these methods often treat
    the query merely as an input feature, without explicitly leveraging
    it to guide either identifier learning or sequence generation. This leaves intent
    guidance weak in both the tokenization and the generation stage,
    causing generated semantic identifiers to drift toward
    high-frequency token patterns rather than the current search intent. In this work, we find that incorporating additional query-related evidence into the framework of end-to-end generative retrieval is important. Based on this finding, we propose IntentGR,
    an intent-enhanced end-to-end framework that strengthens query
    semantics at four complementary stages of the retrieval pipeline.
    Query-Aware Semantic Behavior Representation (QBR) introduces
    query-level behavioral evidence as a new input signal; Query-Semantic
    Alignment (QSA), Long-Tail Query Enrichment (LQE), and
    Query-conditioned Attention Modulation (QAM) then serve as
    integration mechanisms at the optimization, training-distribution, and
    decoding stages to fully exploit this evidence. Experiments on the
    public AOL4PS benchmark and AE, a large-scale industrial e-commerce
    dataset, show that IntentGR improves Recall@10 by 8.11\% and 11.96\%
    over the strongest baseline, with strong gains on
    long-tail queries. These results suggest that enriched query-related input evidence, together with effective integration mechanisms, is beneficial for building intent-aware end-to-end generative search systems.
\end{abstract}

\ccsdesc[500]{Information systems~Recommender systems}

\keywords{Personalized Search, Generative Retrieval, End-to-End Learning}

\maketitle

\section{Introduction}

As a significant task in modern retrieval systems, personalized search~\cite{pitkow2002personalized} aims to return results that are both relevant to user queries and tailored to individual preferences. Different from conventional search~\cite{robertson2009probabilistic, brin1998anatomy, burges2005learning} that primarily relies on query-item relevance, personalized search must incorporate users' historical behaviors to capture individualized preferences. Meanwhile, unlike standard recommendation~\cite{sarwar2001item, he2017neural, covington2016deep, cheng2016wide} which focuses on user-item interactions to estimate user preference, personalized search is required to distinguish a user's needs under different queries. Therefore, an effective personalized search system should establish a semantic connection among query intent, user preference, and target items.

\begin{figure}[t]
  \centering
  \includegraphics[width=\columnwidth]{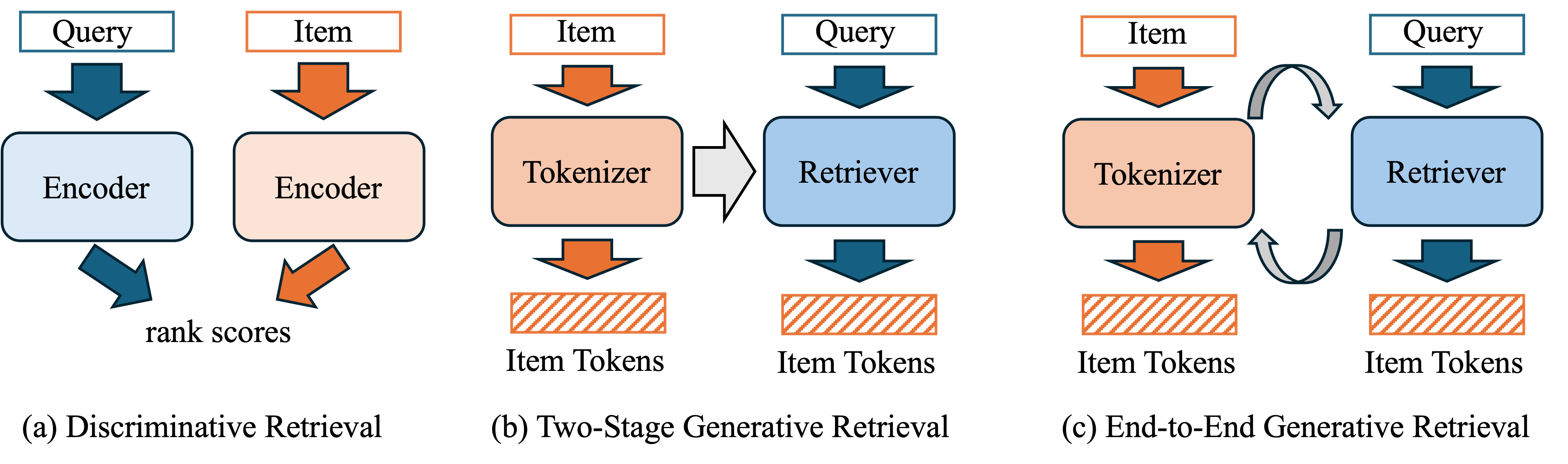}
  \caption{Comparison of three retrieval paradigms. Discriminative retrieval (a) ranks candidates by query-item matching scores; two-stage generative retrieval (b) first constructs item tokens and then trains a retriever; end-to-end generative retrieval (c) jointly optimizes the tokenizer and retriever.}
  \label{fig:gr_paradigms}
\end{figure}

Traditional retrieval approaches typically follow a discriminative
framework~\cite{robertson2009probabilistic, huang2013learning, karpukhin2020dense, covington2016deep, rendle2012bpr, he2017neural}, scoring candidate items for a given query or user and returning the top-ranked results, as shown in Figure~\ref{fig:gr_paradigms} (a).
Retrieval quality thus depends on both the candidate space and the scoring function. Generative retrieval~\cite{de2020autoregressive, tay2022transformer, bevilacqua2022autoregressive} reformulates retrieval as a unified sequence generation task in which the model directly generates item identifiers. Typical generative retrieval methods~\cite{wang2022neural, zhang2025multi, liu2026cat, zhang2024generative, rajput2023recommender} follow a two-stage
pipeline, as illustrated in Figure~\ref{fig:gr_paradigms} (b), which first constructs target identifiers offline from document content, neural indexing models, or item tokenizers, and then trains a retriever to generate these identifier sequences given a query or user context. Fixing the identifiers before retriever training decouples identifier construction from generation, which can leave the resulting identifiers poorly aligned with the downstream retriever's generation objective.
As a consequence, the downstream retriever is inherently constrained by how well the learned identifiers preserve item semantics. To bridge the gap, the end-to-end generative retrieval~\cite{liu2025generative} jointly learns item tokenization and target SID prediction within a unified framework, as displayed in Figure~\ref{fig:gr_paradigms} (c). This enables SIDs to be optimized toward the downstream generation objective, thus becoming more compatible with the generative retriever.

Although end-to-end generative retrieval effectively reduces the mismatch between item tokenization and SID generation, existing designs do not fully satisfy the requirements of personalized search. Recommendation-oriented end-to-end GR methods like ETEGRec~\cite{liu2025generative} mainly focus on user-item interactions and target item generation, leaving query semantics insufficiently explored in the end-to-end workflow. Search-oriented generative methods such as ASI++~\cite{liu2024asi++} move one step further by encoding query semantics into the learned identifier space, yet they treat the query merely as an input signal for aligning and balancing ID assignments, and disregard user behavioral histories entirely.

These gaps motivate a more focused question: \emph{how can end-to-end generative retrieval for personalized search more effectively incorporate and exploit query-related evidence?} In this work, we observe that enriching the model input with query-related behavioral evidence can yield clear performance improvements. This suggests that input-level intent evidence is important for generating semantic identifiers that reflect the current search intent rather than drifting toward high-frequency token patterns. However, simply adding such evidence to the input is not sufficient. To fully exploit the introduced evidence, the generative pipeline also requires appropriate mechanisms to align query semantics with the learned identifier space, balance the training distribution of low-frequency queries, and strengthen query conditioning during decoding.

To this end, we propose IntentGR, an end-to-end framework for personalized generative retrieval in search. While jointly optimizing item tokenization and generative retrieval, IntentGR introduces four key designs to strengthen intent semantics throughout the retrieval process: 
\textbf{1) Query-Aware Semantic Behavior
Representation (QBR)}. It enriches intent evidence at the input level. By adding a query-related behavior sequence, built from earlier interactions
under the same query, QBR places query-level evidence and user-level preference in a single encoder context, helping the model locate the historical interests relevant to the current
query.
\textbf{2) Query-Semantic Alignment (QSA)}. QSA introduces query semantics as an explicit supervisory signal into the joint optimization of item tokenization and generative retrieval. By aligning query representations in generative retrieval with reconstructed target item representations in item tokenization, QSA provides a shared query-item semantic objective for both components, rather than optimizing them only around item reconstruction and SID prediction.
\textbf{3) Long-Tail Query Enrichment (LQE)}. To mitigate head-query dominance, IntentGR enriches long-tail query data by increasing their occurrence and semantic diversity. This design provides more balanced query-level supervision, allowing long-tail intents to play a more active role in end-to-end learning rather than being overwhelmed by frequent query patterns.
\textbf{4) Query-conditioned Attention Modulation (QAM)}. QAM strengthens the guidance of query intent during SID generation. It enhances the attention context used by the decoder with query semantics, further directing the decoder toward the intent expressed in the query. In this way, the decoder generates SID tokens under stronger query guidance, leading to SID sequences that better match the search intent.

Together, these designs make query intent a central guiding signal—rather than a single passive input feature—at four complementary points of the end-to-end pipeline: input representation, joint optimization, training distribution, and decoding context. IntentGR provides a systematic way to incorporate query intent across all four stages of end-to-end generative retrieval for personalized search.

We evaluate IntentGR on a public benchmark and a large-scale industrial dataset. It outperforms the strongest baseline on all metrics, improving Recall@10 by \textbf{8.11}\% and \textbf{11.96}\% respectively, which verifies the effectiveness of the four mechanisms across different search scenarios. 

To summarize, the main contributions of this work are as follows: 

\begin{itemize}
\item We find that augmenting the input of end-to-end generative retrieval with additional query-related evidence can bring clear performance improvements for personalized search, especially when historical query-level evidence provides strong clues about target items.

\item We propose IntentGR, an intent-enhanced end-to-end framework that validates this finding through four complementary mechanisms: Query-Aware Semantic Behavior Representation (QBR) introduces query-level behavioral evidence as input, while Query-Semantic Alignment (QSA), Long-Tail Query Enrichment (LQE), and Query-conditioned Attention Modulation (QAM) serve as integration mechanisms at the optimization, training distribution, and decoding stages that enable the generative pipeline to fully exploit this evidence—lifting performance beyond what either raw behavioral retrieval or shallow query incorporation achieves.
\item Experiments on the public AOL4PS benchmark and a large-scale industrial dataset show consistent improvements, especially on long-tail queries, indicating that enriched query-related input evidence is valuable and that proper integration mechanisms can further improve its utilization.

\end{itemize}

\section{Related Works}

\subsection{Generative Retrieval (GR)}

Traditional information retrieval matches queries to documents using sparse
lexical models~\cite{salton1988term, robertson2009probabilistic} or dense
neural retrievers~\cite{karpukhin2020dense, xiong2020approximate, gao2022unsupervised}, while generative retrieval (GR) refomulates retrieval as sequence-to-sequence generation.
It autoregressively generates semantic identifiers
(SIDs) conditioned on the query and the user's interaction history. Research on identifier design has explored a variety of representations. DSI~\cite{tay2022transformer} assigns hierarchical numeric identifiers to documents, and NCI~\cite{wang2022neural} builds on this scheme with a prefix-aware decoder and pseudo-query augmentation. SEAL~\cite{bevilacqua2022autoregressive}
uses document n-grams as identifiers, while SEATER~\cite{si2024generative}
organizes item identifiers in a balanced $k$-ary tree and contrastively aligns
user representations with identifier paths. TIGER~\cite{rajput2023recommender}
uses RQ-VAE to learn discrete SIDs directly from item representations.
However, SIDs are typically frozen during retriever training, which prevents them
from co-adapting with the generator and makes SID quality a bottleneck
for retrieval performance.

\subsection{End-to-End SID Learning}
\label{subsec:end_to_end}

Motivated by this bottleneck, recent work jointly optimizes the SID tokenizer and generator, an approach we term \emph{End-to-End SID Learning}. Here, ``end-to-end'' refers specifically to joint parameter optimization of these two components.
For sequential recommendation, ETEGRec~\cite{liu2025generative} couples an RQ-VAE item tokenizer with a generator through two alignment losses, alternating updates before fine-tuning the generator with the tokenizer frozen.
For document retrieval, ASI++~\cite{liu2025towards} jointly optimizes ID assignment and retrieval through differentiable quantization and introduces a criterion to balance ID usage.
GLEN~\cite{lee2023glen} initializes lexical IDs with BM25-scored keywords and refines them using ranking signals and a temperature-annealed softmax.
NOVO~\cite{wang2023novo} jointly refines document n-gram IDs and the retriever until the ID set stabilizes.
These methods leave a gap for personalized search: recommendation methods omit query information, while document-retrieval methods disregard user behaviors and treat the query only as a passive decoding condition.
This motivates a framework that uses query semantics to explicitly guide both item tokenization and SID generation and incorporates user interaction histories as a retrieval signal.

\subsection{Query-aware Generative Search}

Since the query is the primary semantic signal that guides identifier generation in search, existing generative retrieval methods explicitly incorporate query information at different stages of the pipeline and can be broadly grouped into three categories.
1) Query-based data enrichment.
DSI-QG~\cite{zhuang2022bridging} represents documents with pseudo-queries to reduce the indexing-retrieval mismatch in DSI~\cite{tay2022transformer}, while NCI~\cite{wang2022neural} uses DocT5Query-generated queries and document terms as training inputs. CorpusBrain~\cite{chen2022corpusbrain} constructs pseudo query-identifier pairs through three pre-training tasks, and CorpusBrain++~\cite{guo2025corpusbrain++} extends it to dynamic corpora through adapter-based continual pre-training with experience replay.
2) Query-driven objective supervision.
LTRGR~\cite{li2024learning} adds a margin-based ranking loss after training with the generation loss, while RIPOR~\cite{zeng2024scalable} applies ranking supervision to each document identifier prefix.
3) Query-oriented architectural design.
OneSearch-V2~\cite{chen2026onesearch} extends OneSearch~\cite{chen2025onesearch} with a thought-augmented query module that uses LLM-generated chains of thought for training alignment and as additional signals at inference, targeting complex and long-tail queries.
These strategies incorporate query information at progressively deeper levels of the pipeline, from training data and objectives to model architecture.

\section{Methodology}

\begin{figure*}[t]
   \vspace{-10pt}
   \centering
   \includegraphics[width=0.95\textwidth, height=0.4\textwidth]{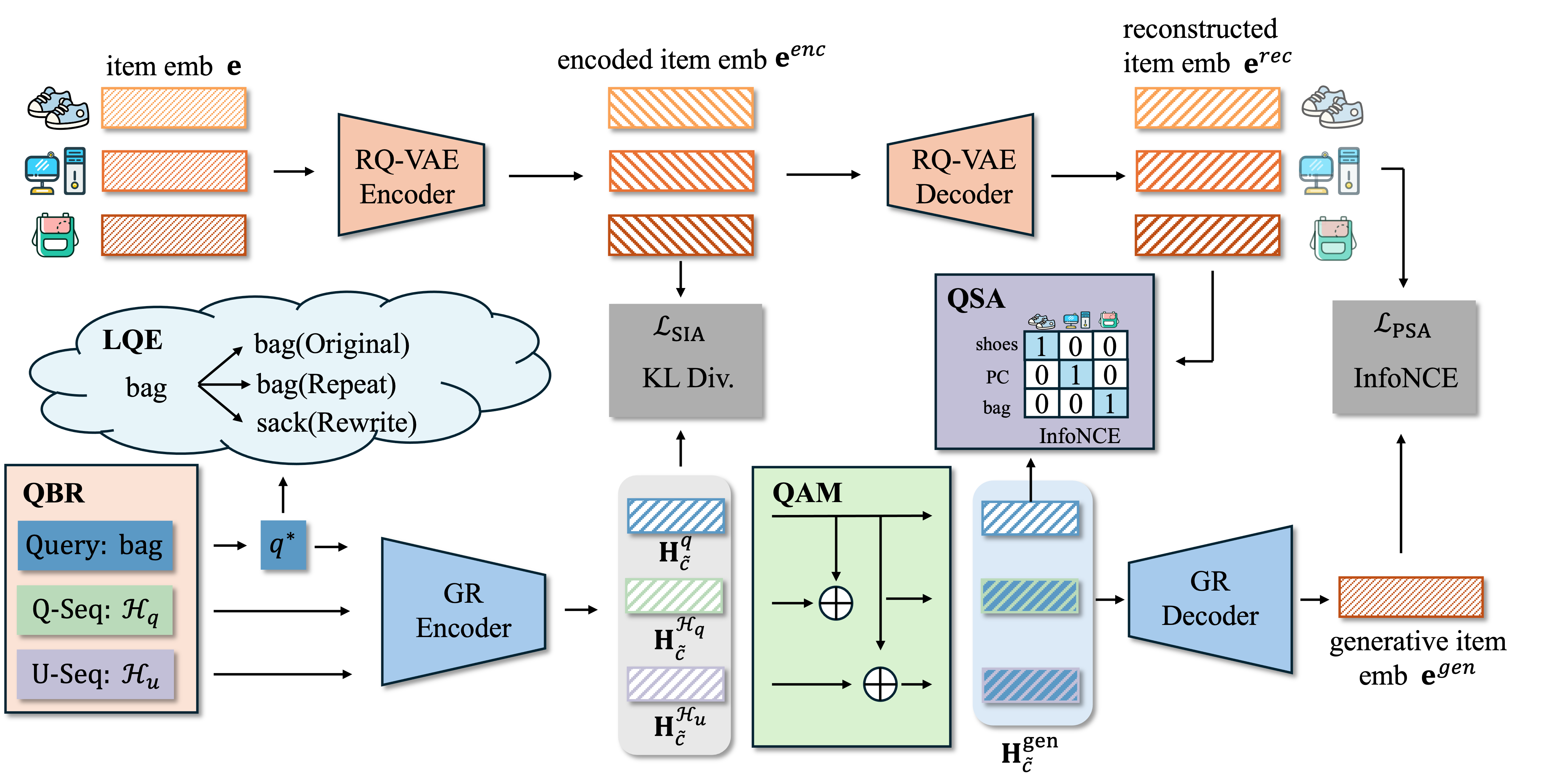}
   \captionsetup{font=small}
   \caption{An overview of IntentGR. IntentGR consists of four key mechanisms: Query-Aware Semantic Behavior
Representation (QBR), Query-Semantic Alignment (QSA), Long-Tail Query Enrichment (LQE), and Query-conditioned Attention Modulation (QAM). These mechanisms enhance intent from complementary perspectives, enabling IntentGR to better capture query intent and personalized preference in search.}
   \label{fig:overview}
   \vspace{-5pt}
\end{figure*}

\textbf{Problem Formulation.}
Let $\mathcal{U}$, $\mathcal{Q}$, and $\mathcal{I}$ denote the user set, query set, and item set, respectively.
For a user-query pair $(u,q)$, where $u \in \mathcal{U}$ and $q \in \mathcal{Q}$, we define the basic personalized search context as $c_{u,q}=(q,\mathcal{H}_u)$, where $\mathcal{H}_u$ denotes the user's historical interaction sequence.
Each training instance is represented as $(c_{u,q}, i^+) \in \mathcal{D}$, where $i^+ \in \mathcal{I}$ is the positive item associated with context $c_{u,q}$, including both clicked and purchased items.
Personalized search aims to retrieve an item that matches both the query intent and the user's personalized preference.
Following generative retrieval, each item $i \in \mathcal{I}$ is represented by a SID sequence $\mathbf{z}_i=(z_i^1,z_i^2,\ldots,z_i^L)$, where $L$ is the SID length and each code $z_i^l$ is selected from a discrete codebook $\mathcal{C}=\{1,\ldots,K\}$.
Since not every code combination corresponds to a valid item, the valid SID space is defined as $\mathcal{Z} \subseteq \mathcal{C}^L$, where each $\mathbf{z}_i \in \mathcal{Z}$ uniquely identifies an item $i \in \mathcal{I}$.
Under this formulation, the model predicts a SID sequence $\hat{\mathbf{z}}=(\hat{z}^1,\hat{z}^2,\ldots,\hat{z}^L) \in \mathcal{Z}$ conditioned on $c_{u,q}$, and the generated SID sequence is then mapped back to the corresponding item $\hat{i} \in \mathcal{I}$ as the retrieval result.

In this section, we first revisit GR baselines in Section~\ref{3.1}. We then present an overview of the proposed IntentGR framework in Figure~\ref{fig:overview} and introduce its four key designs in Sections~\ref{sec:qbr}--\ref{sec:qam}, followed by the training strategy in Section~\ref{sec:strategy}\footnote{AI programming tools were used to assist code implementation. All generated code was manually reviewed, verified, and integrated by the authors.}.

\subsection{Revisit on GR Baselines}
\label{3.1}

Generative retrieval reformulates item retrieval as a sequence generation problem, where the target item is represented and retrieved through a generated identifier sequence~\cite{de2020autoregressive, tay2022transformer, bevilacqua2022autoregressive, wang2022neural}. For personalized item retrieval, the GR framework usually consists of an item tokenizer $T_\phi$ and a generative retriever $R_\theta$~\cite{rajput2023recommender, penha2025semantic, ju2025generative}. The item tokenizer $T_\phi$ maps each item into a discrete SID sequence, while the generative retriever $R_\theta$ generates the target SID sequence conditioned on the input context. GR baselines mainly differ in whether item tokenization is learned separately or jointly with the generative retriever, resulting in two paradigms: two-stage GR and end-to-end GR. In this paper, we build IntentGR upon the end-to-end GR paradigm.

\textbf{Item Tokenizer.}
Given an item $i$ with semantic/collaborative representation $\mathbf{e}_i$, the item tokenizer $T_\phi$ maps it into a discrete SID sequence
$\mathbf{z}_i=(z_i^1,z_i^2,\ldots,z_i^L)$.
The tokenizer can be implemented in different ways, ranging from predefined indexing strategies to offline semantic tokenizers and learnable item tokenizers.
In this paper, we instantiate $T_\phi$ with an RQ-VAE-based tokenizer.
It encodes item representations into latent vectors, quantizes them through multiple codebooks, and reconstructs the item representations to preserve item semantics.
We denote the reconstructed representation of item $i$ as $\mathbf{e}_i^{\mathrm{rec}}$.
The tokenizer-side semantic quantization objective is written as
\begin{equation}
\mathcal{L}_{\mathrm{SQ}}
=
\ell_{\mathrm{SQ}}(\mathbf{e}_i,\mathbf{z}_i,\mathbf{e}_i^{\mathrm{rec}}),
\end{equation}
where $\ell_{\mathrm{SQ}}(\cdot)$ encourages the learned SID codes to retain the semantic information of the original item representation $\mathbf{e}$.
The resulting SID sequence serves as the generation target for the retriever.

\textbf{Generative Retriever.}
Given an input context $c$, the generative retriever $R_\theta$ predicts the SID sequence of the target item $i^+$ in an autoregressive manner.
The retriever may adopt different architectures, such as encoder-decoder models or decoder-only language models.
Its generation probability can be generally written as:
\begin{equation}
p_\theta(\mathbf{z}_{i^+} \mid c)
=
\prod_{l=1}^{L}
p_\theta(z_{i^+}^{l} \mid z_{i^+}^{<l}, c),
\end{equation}
where $\mathbf{z}_{i^+}=T_\phi(\mathbf{e}_{i^+})$ is the SID sequence of the target item and $z_{i^+}^{<l}$ denotes the previously generated SID tokens.
The corresponding generative retrieval loss is defined as the negative log-likelihood of the target SID sequence:
\begin{equation}
\mathcal{L}_{\mathrm{REC}}
=
-\sum_{l=1}^{L}
\log p_\theta(z_{i^+}^{l} \mid z_{i^+}^{<l}, c).
\end{equation}

\textbf{Optimization Paradigms.}
Existing GR baselines mainly follow two optimization paradigms. In two-stage GR~\cite{rajput2023recommender, ju2025generative}, the item tokenizer $T_\phi$ is first trained or constructed independently, and the obtained SID sequences are then fixed as generation targets for training the generative retriever $R_\theta$. This paradigm separates item tokenization from the downstream retrieval generation objective. In contrast, end-to-end GR~\cite{liu2025generative} jointly optimizes $T_\phi$ and $R_\theta$ within a unified framework. Besides the SID prediction loss for the retriever, auxiliary alignment objectives are often introduced to align the encoded input context and generated SID representations with the target item semantics. In this way, end-to-end GR aligns item tokenization with target SID prediction rather than treating them as two isolated stages. In this paper, we build IntentGR upon the end-to-end GR paradigm and further enhance it with intent semantics for personalized search.

\subsection{Query-Aware Semantic Behavior Representation}
\label{sec:qbr}
The basic context for personalized search $c_{u,q}=(q,\mathcal{H}_u)$ contains the query and the user's historical interactions, but it does not provide query-level behavioral evidence that reflects how items are generally interacted with under the same query. Thus IntentGR proposes Query-Aware Semantic Behavior Representation (QBR), which introduces a query-related behavior sequence $\mathcal{H}_q$ and extends the basic context as 
\begin{equation}
\tilde{c}_{u,q}=(q,\mathcal{H}_q,\mathcal{H}_u).
\end{equation} 

Notably, $\mathcal{H}_q$ aggregates clicks from \emph{all users} who issued the same query before the current sample, effectively realizing a form of query-level collaborative filtering within the generative retrieval framework—distinct from conventional user-level CF in that it captures collective intent patterns tied to a specific information need. Given $\tilde{c}_{u,q}$, the query $q$ is transformed into token embeddings $\mathbf{E}_q$, while items in $\mathcal{H}_q$ and $\mathcal{H}_u$ are represented by their SID embeddings, denoted as $\mathbf{E}_{\mathcal{H}_q}$ and $\mathbf{E}_{\mathcal{H}_u}$, respectively. The query-aware encoder input is constructed as an ordered sequence of three semantic segments:
\begin{equation}
\mathbf{X}_{\tilde{c}} =
[\mathbf{E}_q;\mathbf{E}_{\mathcal{H}_q};\mathbf{E}_{\mathcal{H}_u}],
\end{equation}
where the three segments correspond to the query, query-related behaviors, and user history interactions, respectively. When $\mathcal{H}_q$ is unavailable, the input naturally reduces to $[\mathbf{E}_q;  \mathbf{E}_{\mathcal{H}_u}]$.
The constructed input $\mathbf{X}_{\tilde{c}}$ is fed into the encoder of the generative retriever:
\begin{equation}
\mathbf{H}_{\tilde{c}}=\mathrm{Enc}_{\theta}(\mathbf{X}_{\tilde{c}}).
\end{equation}
By organizing these segments within a unified context, IntentGR allows query-level behavioral evidence to interact with user-level preference in the same contextual representation. This helps the model identify which parts of the user's historical interests are more relevant to the current query, resulting in a query-aware representation for personalized SID generation.

\subsection{Query-Semantic Alignment}
\label{sec:qsa}

The query-aware behavior representation introduced above incorporates query semantics into the input context, but it does not directly specify how query intent should guide the joint learning of the retriever and the tokenizer. Therefore, we design Query-Semantic Alignment (QSA), a shared contrastive objective that introduces query-item relevance into the joint optimization.

To instantiate this alignment objective, we use the query-side representation derived from the query-token region of $\mathbf{H}_{\tilde{c}}$ and the reconstructed representation of the target item produced by the tokenizer as the aligned pair. Formally, we define
\begin{equation}
\mathbf{r}_q = g_{q2i}(\mathbf{H}_{\tilde{c}}^{q}), \quad
\mathbf{r}_{i^+} = \mathbf{e}_{i^+}^{\mathrm{rec}},
\end{equation}
where $\mathbf{H}_{\tilde{c}}^{q}$ denotes the encoder outputs corresponding to the query tokens, $g_{q2i}(\cdot)$ maps them into the item semantic space, and $\mathbf{e}_{i^+}^{\mathrm{rec}}$ is the reconstructed target item representation.
This aligned pair connects the retriever and the tokenizer through a shared query-item semantic objective, encouraging both contextual retrieval representations and learned SIDs to be optimized under search-intent supervision.

For each query representation $\mathbf{r}_{q_m}$, its positive sample is the reconstructed representation $\mathbf{r}_{i_m^+}$ of the corresponding target item, while other item representations in the batch are treated as negatives. The query-to-item contrastive loss is defined as
\begin{equation}
\label{eq:q2i}
\mathcal{L}_{q\rightarrow i}
=
-\frac{1}{M}
\sum_{m=1}^{M}
\log
\frac{
\exp(s(\mathbf{r}_{q_m},\mathbf{r}_{i_m^+})/\tau)
}{
\sum_{n=1}^{M}
\exp(s(\mathbf{r}_{q_m},\mathbf{r}_{i_n^+})/\tau)
},
\end{equation}
where $M$ denotes the number of effective sample pairs in the batch, $s(\cdot,\cdot)$ denotes cosine similarity, and $\tau$ is the temperature parameter.
We also apply the contrastive objective in the reverse direction, treating each reconstructed item representation as the anchor and its paired query representation as the positive sample.
The final query-semantic contrastive loss is
\begin{equation}
\label{eq:qi}
\mathcal{L}_{\mathrm{QSA}}
=
\mathcal{L}_{q\rightarrow i}
+
\mathcal{L}_{i\rightarrow q}.
\end{equation}
Through this bidirectional objective, QSA provides intent-level supervision for the end-to-end learning process. The complete training objective will be presented in Section~\ref{sec:strategy}.

\subsection{Long-Tail Query Enrichment}
\label{sec:lqe}

A key requirement of personalized search is to handle diverse query intents, rather than only performing well on frequent search patterns. However, real-world search logs usually exhibit a highly long-tailed query distribution: head queries provide abundant supervision, while many tail queries appear only a few times but still reflect fine-grained user intents. Directly training on such an imbalanced distribution can make the model biased toward head-query patterns and weaken its ability to learn from long-tail intents. To improve the coverage of tail intents in end-to-end generative retrieval, IntentGR devises Long-Tail Query Enrichment (LQE), which enriches long-tail queries at the training data level through repeated sampling and query rewriting.

For each query $q$, we denote its frequency in the training set as $f(q)$. Given a target frequency threshold $\gamma$, we define the long-tail query set as
\begin{equation}
\mathcal{Q}_{tail}=\{q\in\mathcal{Q}\mid f(q)<\gamma\}.
\end{equation}
For each training instance associated with a long-tail query $q\in\mathcal{Q}_{tail}$, IntentGR increases its training occurrence according to a query-level repeat factor:
\begin{equation}
r(q)=
\left\lceil \frac{\gamma}{f(q)} \right\rceil.
\end{equation}
In this way, samples from low-frequency queries are presented more often during training, giving tail queries more opportunities to update the model.

Repeated sampling increases the occurrence of tail queries, but it does not introduce new semantic expressions. Therefore, IntentGR further applies query rewriting to enrich the semantic forms of long-tail queries. Given a valid rewrite $q'$ of a tail query $q$, we construct an additional training sample by replacing the original query text with $q'$ while keeping the associated behavior context and target item unchanged:
\begin{equation}
(\tilde{c}_{u,q}, i^+)
=
((q,\mathcal{H}_q,\mathcal{H}_u), i^+)
\Rightarrow
(\tilde{c}_{u,q}^{\,q'}, i^+)
=
((q',\mathcal{H}_q,\mathcal{H}_u), i^+).
\end{equation}
This produces semantically-related query variants for the same interaction signal, helping the model learn more robust intent representations for tail queries.

As a result, long-tail queries are presented more frequently and with more diverse expressions. We apply this process only to the training data and keep the validation and test sets unchanged to ensure fair evaluation. With LQE balancing the query-level training distribution, IntentGR reduces the dominance of head-query patterns and allows long-tail intents to participate more effectively in end-to-end generative retrieval.

\subsection{Query-Conditioned Attention Modulation}
\label{sec:qam}

After the enhanced context $\tilde{c}_{u,q}$ is encoded into the contextual representation $\mathbf{H}_{\tilde{c}}$, the generative retriever produces the target SID sequence in an autoregressive manner, where the decoder attends to
$\mathbf{H}_{\tilde{c}}$ through cross-attention at every generation step. During autoregressive generation, the influence of the query on the decoder depends on how much cross-attention is allocated to the query positions.
However, the query occupies only a small fraction of the encoded context, while most positions in $\mathbf{H}_{\tilde{c}}$ correspond to behavior items. Consequently, the decoder's attention is easily dominated by item semantics, and the query signal can be diluted during generation. To address this, we propose Query-conditioned Attention Modulation (QAM), which explicitly injects query semantics into the contextual representation that the decoder attends to, steering its cross-attention toward the search intent.

Specifically, let $\mathbf{H}_{\tilde{c}}^{q}$ and $\mathbf{H}_{\tilde{c}}^{\neg q}$ denote the representations of the query tokens and the remaining context tokens in $\mathbf{H}_{\tilde{c}}$, respectively. QAM first aggregates the query representation $\mathbf{H}_{\tilde{c}}^{q}$ into a global query vector:
\begin{equation}
\mathbf{v}_q = \mathrm{Pool}(\mathbf{H}_{\tilde{c}}^{q}).
\end{equation}
Then QAM expands $\mathbf{v}_q$ to the sequence length of the non-query states and denote the expanded query representation as $\mathbf{V}_q$.
Afterwards, QAM fuses $\mathbf{V}_q$ into every non-query position and constructs query-aware generation context as
\begin{equation}
\label{eq:qam}
\mathbf{H}_{\tilde{c}}^{\mathrm{gen}}
=
\left[
\mathbf{H}_{\tilde{c}}^{q};
\alpha \mathbf{H}_{\tilde{c}}^{\neg q}
+
(1-\alpha)\mathbf{V}_q
\right],
\end{equation}
where $\alpha \in (0,1)$ is a hyperparameter that adaptively balances the original contextual semantics and the injected query semantics.

The decoder then performs cross-attention over $\mathbf{H}^{\mathrm{gen}}$ instead of $\mathbf{H}_{\tilde{c}}$. In this way, every position in the
context carries query-conditioned information, so the decoder receives query semantics regardless of where its attention is distributed, and the generated SID sequence follows the search intent more faithfully.

\subsection{Training Strategy}
\label{sec:strategy}

\begin{figure}[t]
  \centering
  \includegraphics[width=\columnwidth]{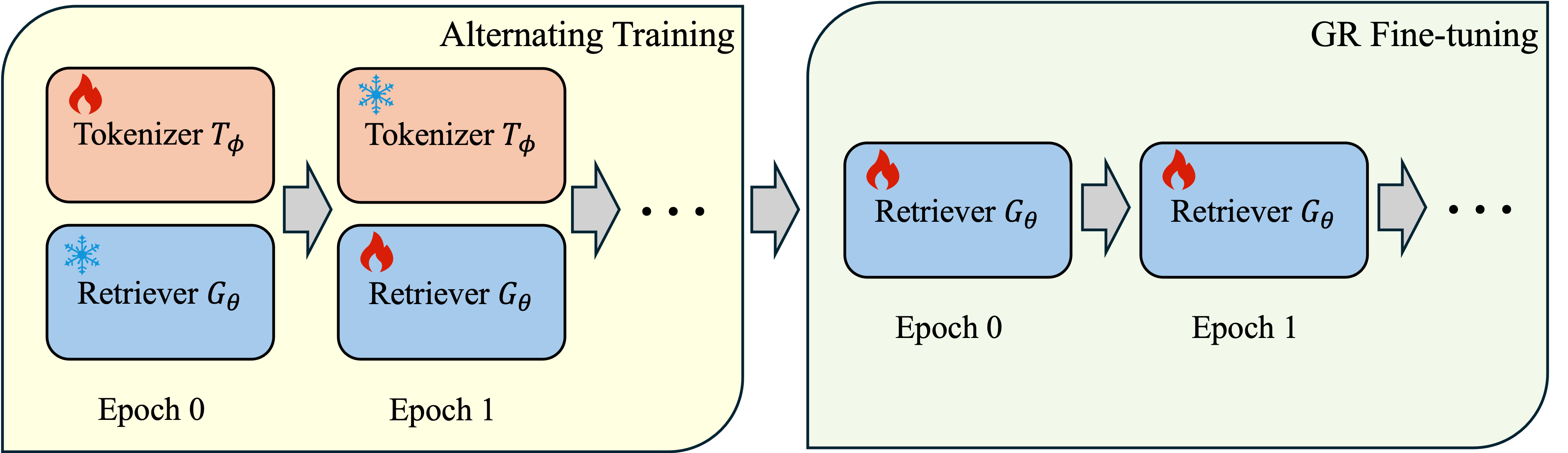}
  \caption{The training strategy of IntentGR, consisting of alternating training, where $T_\phi$ and $R_\theta$ are updated in turn, and GR fine-tuning, where $T_\phi$ is frozen and $R_\theta$ is further optimized.}
  \vspace{-1.2em}
  \label{fig:training_strategy}
\end{figure}

Following the end-to-end paradigm in Section~\ref{3.1}, IntentGR first performs alternating training, which iterates between a tokenizer optimization step and a retriever optimization step: the former updates $T_\phi$ with $R_\theta$ fixed, and the latter updates $R_\theta$ with
$T_\phi$ fixed, allowing the SID space and the retriever to adapt to each other. After the alternating training converges, $T_\phi$ is frozen and $R_\theta$ is further optimized on the stabilized SID space, referred to as GR fine-tuning.

\subsubsection{Alternating Training}
To further bridge item tokenization and generative retrieval, we adopt two auxiliary alignment losses~\cite{liu2025generative}: the Sequence-Item Alignment loss $\mathcal{L}_{\mathrm{SIA}}$, which matches the code assignment distributions of the context representation and the target item embedding $\mathbf{e}^{enc}$ via a symmetric KL divergence, and the Preference-Semantic Alignment loss $\mathcal{L}_{\mathrm{PSA}}$, which aligns the GR decoder's item representation $\mathbf{e}^{gen}$ with the reconstructed item embedding $\mathbf{e}^{rec}$ via InfoNCE. 

Specifically, the context representation is obtained by pooling the encoder output, i.e., $\mathbf{z}^{\tilde{c}} =
\operatorname{MLP}(\operatorname{Pool}(\mathbf{H}_{\tilde{c}}))$, and $\mathcal{L}_{\mathrm{SIA}}$ is defined as
\begin{equation}
\mathcal{L}_{\mathrm{SIA}}
= \sum_{l=1}^{L}
\Big(
D_{\mathrm{KL}}\big(P^{l}_{\mathbf{e}^{enc}} \,\|\, P^{l}_{\mathbf{z}^{\tilde{c}}}\big)
+ D_{\mathrm{KL}}\big(P^{l}_{\mathbf{z}^{\tilde{c}}} \,\|\, P^{l}_{\mathbf{e}^{enc}}\big)
\Big),
\label{eq:sia}
\end{equation}
where $P^{l}_{\mathbf{x}}$ denotes the token distribution at the $l$-th level produced by the tokenizer for input $\mathbf{x}$, and $D_{\mathrm{KL}}(\cdot\|\cdot)$ is the Kullback-Leibler divergence. 
On the generation side, each generated item representation
$\mathbf{e}^{gen}_m$ is contrasted against the reconstructed embeddings in
the batch, with that of its target item, $\mathbf{e}^{rec}_m$, as the
positive sample:
\begin{equation}
\mathcal{L}_{g \to r}
= -\frac{1}{M} \sum_{m=1}^{M}
\log \frac{\exp\big(s(\mathbf{e}^{gen}_m, \mathbf{e}^{rec}_m)/\tau\big)}
          {\sum_{n=1}^{M} \exp\big(s(\mathbf{e}^{gen}_m, \mathbf{e}^{rec}_n)/\tau\big)},
\label{eq:psa_g2r}
\end{equation}
where $M$, $s(\cdot,\cdot)$, and $\tau$ are as defined in
Eq.~\eqref{eq:q2i}. Combined with the reverse direction, we derive $\mathcal{L}_{\mathrm{PSA}}$ as
\begin{equation}
\mathcal{L}_{\mathrm{PSA}} = \mathcal{L}_{g \to r} + \mathcal{L}_{r \to g}.
\label{eq:psa}
\end{equation}

The two alternating steps share the two alignment losses and the query-semantic contrastive loss $\mathcal{L}_{\mathrm{QSA}}$, and differ in their primary objectives——the tokenizer step focuses on the quantization loss $\mathcal{L}_{\mathrm{SQ}}$, whereas the retriever step centers on the SID prediction loss $\mathcal{L}_{\mathrm{REC}}$:
\begin{equation}
\begin{aligned}
\label{loss_it_gr}
\mathcal{L}_{\mathrm{IT}}
&=
\mathcal{L}_{\mathrm{SQ}}
+
\mathcal{L}_{\mathrm{SIA}}
+
\mathcal{L}_{\mathrm{PSA}}
+
\eta\mathcal{L}_{\mathrm{QSA}},\\
\mathcal{L}_{\mathrm{GR}}
&=
\mathcal{L}_{\mathrm{REC}}
+
\mathcal{L}_{\mathrm{SIA}}
+
\mathcal{L}_{\mathrm{PSA}}
+
\eta\mathcal{L}_{\mathrm{QSA}},
\end{aligned}
\end{equation}
where $\eta$ is the weighting coefficient for the query-semantic contrastive objective. After each tokenizer optimization step, item SID sequences are refreshed with the updated tokenizer and used in the following retriever optimization step.

\subsubsection{GR Fine-tuning}
After alternating training, we freeze $T_\phi$ and fine-tune $R_\theta$ on the learned SID space. It should be noted that the auxiliary alignment losses are disabled during fine-tuning, and IntentGR optimizes the retriever with the generative retrieval loss and the query-semantic contrastive loss:
\begin{equation}
\label{loss_ft}
\mathcal{L}_{\mathrm{FT}}
=
\mathcal{L}_{\mathrm{REC}}
+
\eta\mathcal{L}_{\mathrm{QSA}}.
\end{equation}
In this way, the retriever is fully adapted to the stabilized SID space while $\mathcal{L}_{\mathrm{QSA}}$ continues to align the generated SIDs with the query semantics.

\section{Experiments}

\subsection{Experimental Setup}

\subsubsection{Datasets}
We conduct experiments on two personalized search datasets: AOL4PS~\cite{aol4ps} and AE. AOL4PS is a public benchmark for personalized search constructed from the AOL query log, where each sample consists of a query, the user's historical behaviors, and the target document. We adopt the official dataset split. Samples are constructed per user in chronological order: each labeled interaction serves as the target item, with all preceding interactions forming the behavior sequence. AE is a real-world e-commerce search dataset collected from the online search logs of AliExpress. It contains one week of user click behaviors. We use all interactions before the last hour of the final day for training, and use the last hour of the final day for evaluation. Compared with AOL4PS, AE is collected from an industrial search scenario with a much larger item corpus. User queries on AE are typically short and colloquial, whereas product titles are often keyword-stacked by sellers, resulting in a pronounced query-item lexical gap. These properties make AE a more challenging large-scale real-world setting for evaluating IntentGR. For both datasets, we construct a query-level behavior sequence for each sample using the most frequently clicked items under the same query, counting only clicks that occur strictly before the current sample to avoid future information leakage. Dataset statistics are summarized in Table~\ref{tab:datasets}.

\subsubsection{Baselines}
We compare IntentGR with representative methods spanning three groups: \textbf{discriminative retrieval}—BM25~\cite{robertson2009probabilistic} (lexical term matching), TwoTower~\cite{yang2020mixed} (dual-encoder embedding similarity), and SASRec~\cite{kang2018self} (self-attentive sequential scoring without query access); \textbf{generative retrieval}—DSI~\cite{tay2022transformer} (clustering-based semantic identifiers), TIGER~\cite{rajput2023recommender} (two-stage RQ-VAE tokenizer + generator), and ETEGRec~\cite{liu2025generative} (end-to-end joint optimization of tokenization and SID prediction); and \textbf{a controlled baseline}, ETEGRec-Q, which adapts ETEGRec for search by incorporating the query into the input context. ETEGRec-Q shares the same backbone, RQ-VAE tokenizer, alternating training paradigm, and data pipeline with IntentGR, differing \emph{only} in the four proposed intent mechanisms—isolating their contribution from confounds such as identifier type or model capacity.

\begin{table}[t]
  \centering
  \caption{Statistics of the two personalized search datasets.}
  \label{tab:datasets}
  \vspace{-0.8em}
  \small
  \setlength{\tabcolsep}{5pt}
  \renewcommand{\arraystretch}{1.08}
  \begin{tabular}{|l|rrrr|}
    \hline
    \rowcolor{mygray}
    Dataset & Users & Queries & Items & Samples \\
    \hline\hline
    AOL4PS & 12,907 & 115,865 & 72,548 & 326,146 \\
    AE & 11,297,940 & 13,433,930 & 9,159,889 & 48,339,816 \\
    \hline
  \end{tabular}
  \vspace{-5pt}
\end{table}

\subsubsection{Evaluation Metrics}
We follow the common evaluation protocol of generative retrieval and measure all experimental results with three metrics: 1) \textbf{Recall@K(R@K)}: This metric evaluates whether the ground-truth item appears in the top-K retrieved results, reflecting the overall retrieval coverage. 2) \textbf{NDCG@K(N@K)}: This metric further considers the ranking position of the ground-truth item within the top-$K$ list, assigning higher scores when the target item is ranked closer to the top. 3) \textbf{MRR@K(M@K)}: This metric measures the reciprocal rank of the ground-truth item, and is more sensitive to whether the target item can be retrieved at the very top positions. For all metrics, we report results with K $ \in \{10,20\}$ and average the scores over all test instances.

\subsubsection{Implementation Details}

We implemented IntentGR with PyTorch and built the generative retriever on the Transformers library. Experiments were conducted on 4$\times$NVIDIA Tesla V100 GPUs for AOL4PS and 8$\times$NVIDIA A100 GPUs for AE. For AOL4PS, the generative retriever was initialized from a pretrained T5-small checkpoint, and the RQ-VAE-based item tokenizer used a SID length of 3 with a codebook size of 256. Item semantic embeddings, used to train the RQ-VAE tokenizer, were obtained by encoding each item's title and URL with BGE-base-en-v1.5~\cite{bge_embedding}. We use the AdamW optimizer with a cosine learning-rate schedule and an effective batch size of 2,048. For QBR, the maximum lengths of query text, query-related behavior sequence, and historical interaction sequence were 32, 50, and 50. For QSA, the contrastive loss coefficient was $\eta=4\times10^{-4}$ in Eq.~\ref{loss_it_gr} and Eq.~\ref{loss_ft}; for LQE, the query frequency threshold was 2, with at most one rewritten query per eligible long-tail query; for QAM, the fusion coefficient was $\alpha=0.8$ in Eq.~\ref{eq:qam}. The effects of $\eta$ and $\alpha$ are further analyzed in Section~\ref{sec:hyper_param_analysis}. For AE, we kept the same framework and evaluation protocol, and adjusted training hyper-parameters to accommodate its larger scale. Besides, we use item collaborative embeddings, learned by a Transformer on query-related behavior sequence, to train the RQ-VAE tokenizer.

\begin{table*}[!t]
  \centering
  \caption{Evaluation results (\%) on AOL4PS and AE datasets. The best results are highlighted in bold font. The standard deviation of the reported results is below 0.1\% across all metrics.
 }
  \vspace{-0.8em}
  \small
  \setlength{\tabcolsep}{4pt}
  \renewcommand{\arraystretch}{1.08}
  \begin{tabular}{|r|l|l||cccccc|cccccc|}
    \hline
    \rowcolor{mygray}
    & & & \multicolumn{6}{c|}{AOL4PS} & \multicolumn{6}{c|}{AE} \\
    \rowcolor{mygray}
    \multirow{-2}[-1]{*}{Method} & \multirow{-2}[-1]{*}{Model} &\multirow{-2}[-1]{*}{Venue}
    & \multicolumn{1}{c}{R@10}
    & \multicolumn{1}{c}{R@20}
    & \multicolumn{1}{c}{N@10}
    & \multicolumn{1}{c}{N@20}
    & \multicolumn{1}{c}{M@10}
    & \multicolumn{1}{c|}{M@20}
    & \multicolumn{1}{c}{R@10}
    & \multicolumn{1}{c}{R@20}
    & \multicolumn{1}{c}{N@10}
    & \multicolumn{1}{c}{N@20}
    & \multicolumn{1}{c}{M@10}
    & \multicolumn{1}{c|}{M@20} \\
    \hline\hline

    \multirow{3}{*}{Discriminative} & BM25  & FnTIR'09
      & 60.00 & 67.14 & 45.76 & 47.56 & 41.23 & 41.72
      & 10.32 & 13.94 & 5.70 & 6.61 & 4.29 & 4.53 \\

    & TwoTower & WWW'20
      & 17.63 & 21.68 & 12.82 & 13.89 & 11.39 & 11.71
      & 3.80 & 5.39 & 2.12 & 2.52 & 1.60 & 1.71 \\

     & SASRec & ICDM'18
      & 40.79 & 44.89 & 33.14 & 34.18 & 30.75 & 31.04
      & 1.88 & 2.65 & 1.20 & 1.39 & 0.99 & 1.04 \\
      
    \hline

    \multirow{5}{*}{Generative} & DSI & NeurIPS'22
      & 22.98 & 23.39 & 21.71 & 21.82 & 21.30 & 21.33
      & 6.36 & 7.09 & 4.62 & 4.80 & 4.08 & 4.13 \\

    & TIGER & NeurIPS'23
      & 43.34 & 45.56 & 36.11 & 36.67 & 33.78 & 33.93
     & 19.24 & 20.84 & 12.40 & 12.81 & 10.23 & 10.34 \\

    & ETEGRec & SIGIR'25
      & 47.03 & 49.52 & 37.59 & 38.23 & 34.58 & 34.76
      & 19.91 & 20.95 & 12.73 & 13.00 & 10.44 & 10.52 \\
      
    & ETEGRec-Q & --
      & 69.88 & 71.00 & 66.35 & 66.64 & 65.21 & 65.29
      & 22.40 & 25.50 & 15.88 & 16.67 & 13.83 & 14.05\\

    & IntentGR & --
      & \textbf{77.99} & \textbf{78.81} & \textbf{74.33} & \textbf{74.54} & \textbf{73.13} & \textbf{73.19}
      & \textbf{34.36} & \textbf{37.19} & \textbf{22.38} & \textbf{23.12} & 
      \textbf{18.63} & \textbf{18.83}\\

    \hline
  \end{tabular}
  \label{tab:sota}
  \vspace{-5pt}
\end{table*}

\begin{table}[t]
  \centering
  \caption{Long-tail retrieval results (\%) on AOL4PS and AE.}
  \label{tab:long_tail_query}
  \vspace{-0.8em}
  \small
  \setlength{\tabcolsep}{3pt}
  \renewcommand{\arraystretch}{1.08}
  \begin{adjustbox}{max width=\columnwidth}
  \begin{tabular}{|c|c|c|c|c|ccc|}
    \hline
    \rowcolor{mygray}
    Dataset & Bucket & Samples(\%) & Queries(\%) & Model & R@10 & N@10 & M@10 \\
    \hline\hline

    \multirow{6}{*}{AOL4PS}
      & \multirow{2}{*}{Low}
      & \multirow{2}{*}{50.34}
      & \multirow{2}{*}{87.03}
      & ETEGRec-Q & 41.47 & 36.70 & 35.18 \\
      & & & & IntentGR & \textbf{57.32} & \textbf{52.21} & \textbf{50.57} \\
    \cline{2-8}

      & \multirow{2}{*}{Medium}
      & \multirow{2}{*}{6.11}
      & \multirow{2}{*}{5.57}
      & ETEGRec-Q & 95.49 & 92.77 & 91.86 \\
      & & & & IntentGR & \textbf{97.58} & \textbf{94.73} & \textbf{93.76} \\
    \cline{2-8}

      & \multirow{2}{*}{High}
      & \multirow{2}{*}{43.56}
      & \multirow{2}{*}{7.40}
      & ETEGRec-Q & 99.11 & 96.92 & 96.17 \\
      & & & & IntentGR & \textbf{99.13} & \textbf{97.03} & \textbf{96.32} \\
    \hline

    \multirow{6}{*}{AE}
      & \multirow{2}{*}{Low}
      & \multirow{2}{*}{57.29}
      & \multirow{2}{*}{58.94}
      & ETEGRec-Q & 17.70 & 12.67 & 11.08 \\
      & & & & IntentGR & \textbf{29.93} & \textbf{19.27} & \textbf{15.93} \\
    \cline{2-8}

      & \multirow{2}{*}{Medium}
      & \multirow{2}{*}{13.48}
      & \multirow{2}{*}{13.73}
      & ETEGRec-Q & 24.43 & 16.63 & 14.19 \\
      & & & & IntentGR & \textbf{41.31} & \textbf{26.16} & \textbf{21.43} \\
    \cline{2-8}

      & \multirow{2}{*}{High}
      & \multirow{2}{*}{29.23}
      & \multirow{2}{*}{27.33}
      & ETEGRec-Q & 30.69 & 21.82 & 19.05 \\
      & & & & IntentGR & \textbf{39.76} & \textbf{26.69} & \textbf{22.59} \\
    \hline
  \end{tabular}
  \end{adjustbox}
  \vspace{-5pt}
\end{table}

\subsection{Overall Performance}

Table~\ref{tab:sota} reports the overall performance on AOL4PS and AE. We compare IntentGR with representative baselines from two paradigms: discriminative retrieval methods and generative retrieval methods.

\noindent\textbf{1) Results on AOL4PS.}
Among the discriminative baselines, BM25 achieves the strongest performance (60.00\% Recall@10), benefiting from the high lexical overlap between queries and document titles in web search logs, while TwoTower is less effective due to the difficulty of learning dense matching from limited data. SASRec, which models only the behavior sequence without access to the query, falls clearly behind BM25 (40.79\% vs. 60.00\% Recall@10), providing the first evidence that query information is indispensable for this task.  Among the generative methods, DSI performs the weakest, as its clustering-based identifiers have limited semantic expressiveness; TIGER and ETEGRec perform substantially better by learning semantic identifiers through RQ-VAE, and ETEGRec further benefits from the end-to-end optimization paradigm. Nevertheless, all query-agnostic generative methods still trail BM25, indicating that improved item tokenization alone cannot compensate for the absence of query intent. Once the query is incorporated, ETEGRec-Q boosts Recall@10 from 47.03\% to 69.88\% and surpasses all baselines by a large margin, confirming that query intent is the decisive factor in personalized generative retrieval. However, ETEGRec-Q merely feeds the query as an additional encoder input, leaving its interaction with item tokenization and SID generation shallow. In contrast, IntentGR achieves the best performance across all metrics, raising \textbf{Recall@10} to \textbf{77.99\%} and \textbf{NDCG@10} to \textbf{74.33\%}, an improvement of over 8\% against ETEGRec-Q. Beyond retrieving more target items, IntentGR also ranks them closer to the top, as reflected by the gains on NDCG and MRR. These improvements demonstrate that the consistent enhancement of intent throughout the end-to-end GR process yields substantial benefits beyond simply feeding the query into the model.

\noindent\textbf{2) Results on AE}.
On AE, the absolute performance of all methods drops substantially, reflecting the difficulty of retrieval over a much larger industrial item corpus. Notably, BM25, a strong discriminative baseline on AOL4PS, degrades sharply (10.32\% Recall@10), as the pronounced lexical gap between queries and product titles undermines exact term matching; TwoTower and SASRec perform even worse, since matching or behavior modeling alone can hardly locate the target within such a vast item space. Among the generative methods, DSI with its clustering-based identifiers still falls behind BM25, whereas TIGER and ETEGRec built on RQ-VAE-based SIDs surpass it by a large margin, indicating that the quality of item tokenization becomes decisive when lexical matching fails. More importantly, ETEGRec-Q improves over ETEGRec by only 2.49\% on Recall@10, far less than the gain it achieves on AOL4PS, suggesting that shallow query concatenation cannot fully exploit the query intent. In contrast, IntentGR outperforms ETEGRec-Q by 11.96\% on \textbf{Recall@10} (22.40\%$\rightarrow$\textbf{34.36\%}), together with consistent gains on NDCG and MRR. These results confirm that the benefits of intent enhancement generalize from the public benchmark to large-scale real-world personalized search scenarios.

\subsection{Analysis on Long-Tail Queries}
\label{sec:long_tail_query_analysis}

To further evaluate the adaptability of IntentGR to queries of different popularity, we analyze retrieval performance under different query-frequency groups. Specifically, test samples are divided into three buckets according to the frequency of their queries in the training set: Low (0--5), Medium (6--10), and High (11+) for AOL4PS, where the thresholds are adjusted to 0--5, 6--30, and 31+ for AE to accommodate its larger scale. As shown in Table~\ref{tab:long_tail_query}, the Low bucket accounts for the vast majority of unique queries on both datasets (87.03\% on AOL4PS and 58.94\% on AE), indicating that long-tail queries constitute a substantial portion of query diversity in personalized search.

On AOL4PS, ETEGRec-Q performs much worse on long-tail queries: its Recall@10 drops from 99.11\% on the High bucket to 41.47\% on the Low bucket, showing that even with query information incorporated, the baseline still struggles when query-level observations are sparse. IntentGR brings consistent improvements across all buckets, with the largest gains on the Low bucket, where Recall@10, NDCG@10, and MRR@10 are improved by over 15\% each. Meanwhile, IntentGR remains comparable or slightly better on the Medium and High buckets, where ETEGRec-Q is already near saturation, indicating that the long-tail gains are not obtained at the cost of head-query performance. 
On AE, where no bucket reaches saturation, IntentGR delivers large improvements across all query-frequency groups, raising Recall@10 by more than 9\% on every bucket. In particular, the substantial gain on long-tail queries persists in this industrial scenario, confirming that the improvement observed on AOL4PS generalizes to large-scale real-world search scenarios. 
Overall, these results suggest that intent enhancement helps the end-to-end GR framework exploit sparse query evidence, yielding substantial long-tail gains without sacrificing head-query performance.

\subsection{Ablation Study}

\subsubsection{Contributions of Intent Enhancement Designs}

As shown in Table~\ref{tab:ablation_designs}, the ablation results show that introducing query-level behavioral evidence through QBR brings a clear performance gain, as it provides the model with additional intent information that is absent from the baseline. Building on this enriched input, QSA, LQE, and QAM yield further improvements by strengthening how this evidence is optimized, balanced, and utilized during generation. Specifically, QSA encourages the learned semantic identifier space to remain aligned with query semantics, LQE increases the exposure of low-frequency queries during training, and QAM reinforces query conditioning in the decoding process. As shown in Table~\ref{tab:long_tail_query}, these mechanisms produce more noticeable benefits in challenging scenarios, such as long-tail queries, where intent signals are sparse. This suggests that QSA, LQE, and QAM serve as important complementary mechanisms that help the model more effectively exploit the behavioral evidence introduced by QBR.

The QBR standalone experiment makes this integration role explicit. QBR as a non-parametric retrieval baseline—directly returning items from the query's historical click sequence—achieves 69.36\% Recall@10, comparable to the generative baseline ETEGRec-Q (69.88\%). Yet IntentGR reaches 77.99\%, surpassing QBR standalone by 8.63\%. This gap is precisely the value delivered by the generative framework equipped with QSA, LQE, and QAM: they enable the model to abstract and generalize beyond raw behavioral co-occurrence, transforming a popularity-level signal into a semantically grounded generation target. Without these integration mechanisms, the behavioral evidence remains at its ``raw retrieval'' ceiling; with them, the same evidence is amplified through learned tokenization alignment, balanced query coverage, and intent-conditioned decoding.

To diagnose \emph{where} the generative pipeline creates value, we partition test samples by whether the target item appears in the query's behavior history (Table~\ref{tab:no_match_query}). The Match bucket is most informative: ETEGRec-Q achieves only 89.11\% Recall@10 despite the target being present in the behavioral context, indicating frequent SID generation failures when query intent is weakly integrated. IntentGR reaches 99.72\% (+10.61\%), demonstrating that QBR, together with the integration mechanisms, substantially reduces this generation failure mode. In the No Match bucket (52.91\% of queries), where no behavioral evidence exists for the target, IntentGR still yields modest gains over ETEGRec-Q (+2.44\% on AOL4PS; +1.22\% on AE). This is expected: both models face the same information-theoretic constraint of relying solely on query text, and the improvement reflects residual benefits from better-calibrated SID spaces (via QSA), decoding guidance (via QAM) and enriched tail-query training (via LQE) rather than behavioral memorization. The user-activity analysis (Table~\ref{tab:inactive_user}) corroborates this interpretation: inactive users—including those entirely unseen during training on AE—benefit comparably to active ones, confirming that gains are grounded in shared query-side patterns rather than user-level memorization.

\begin{table}[t]
  \centering
  \caption{Ablation study of IntentGR on AOL4PS.}
  \label{tab:ablation_designs}
  \vspace{-0.8em}
  \small
  \setlength{\tabcolsep}{7pt}
  \renewcommand{\arraystretch}{1.08}
  \begin{tabular}{|l||c|c|c|}
    \hline
    \rowcolor{mygray}
    Method & R@10 & N@10 & M@10 \\
    \hline\hline
     ETEGRec-Q & 69.88 & 66.35 & 65.21 \\
     QBR & 69.36 & 68.41 & 68.09 \\
     ETEGRec-Q+QBR & 77.44 & 73.87 & 72.70 \\
    ETEGRec-Q+QBR+QSA & 77.55 & 73.98 & 72.81 \\
    ETEGRec-Q+QBR+QSA+LQE & 77.84 & 74.24 & 73.06 \\
    ETEGRec-Q+QBR+QSA+LQE+QAM
      & \textbf{77.99} & \textbf{74.33} & \textbf{73.13} \\
    \hline
  \end{tabular}
  \vspace{-5pt}
\end{table}

\begin{table}[t]
  \centering
  \caption{No match query sequences retrieval results (\%) on AOL4PS and AE.}
  \label{tab:no_match_query}
  \vspace{-0.8em}
  \small
  \setlength{\tabcolsep}{3pt}
  \renewcommand{\arraystretch}{1.08}
  \begin{adjustbox}{max width=\columnwidth}
  \begin{tabular}{|c|c|c|c|c|ccc|}
    \hline
    \rowcolor{mygray}
    Dataset & Bucket & Samples(\%) & Queries(\%) & Model & R@10 & N@10 & M@10 \\
    \hline\hline

    \multirow{4}{*}{AOL4PS}
      & \multirow{2}{*}{No Match}
      & \multirow{2}{*}{30.63}
      & \multirow{2}{*}{52.91}
      & ETEGRec-Q & 26.32 & 20.12 & 18.12 \\
      & & & & IntentGR & \textbf{28.76} & \textbf{20.96} & \textbf{18.45} \\
    \cline{2-8}

      & \multirow{2}{*}{Match}
      & \multirow{2}{*}{69.37}
      & \multirow{2}{*}{47.09}
      & ETEGRec-Q & 89.11 & 86.77 & 86.00 \\
      & & & & IntentGR & \textbf{99.72} & \textbf{97.90} & \textbf{97.28} \\
    \hline

    \multirow{4}{*}{AE}
      & \multirow{2}{*}{No Match}
      & \multirow{2}{*}{75.78}
      & \multirow{2}{*}{75.92}
      & ETEGRec-Q & 17.93 & 12.31 & 10.17 \\
      & & & & IntentGR & \textbf{19.15} & \textbf{12.44} & \textbf{10.73} \\
    \cline{2-8}

      & \multirow{2}{*}{Match}
      & \multirow{2}{*}{24.22}
      & \multirow{2}{*}{24.08}
      & ETEGRec-Q & 36.41 & 26.65 & 23.55 \\
      & & & & IntentGR & \textbf{81.84} & \textbf{53.83} & \textbf{45.04} \\

    \hline
  \end{tabular}
  \end{adjustbox}
  \vspace{-5pt}
\end{table}

\begin{table}[t]
  \centering
  \caption{Inactive user retrieval results (\%) on AOL4PS and AE.}
  \label{tab:inactive_user}
  \vspace{-0.8em}
  \small
  \setlength{\tabcolsep}{3pt}
  \renewcommand{\arraystretch}{1.08}
  \begin{adjustbox}{max width=\columnwidth}
  \begin{tabular}{|c|c|c|c|c|ccc|}
    \hline
    \rowcolor{mygray}
    Dataset & Bucket & Samples(\%) & Users(\%) & Model & R@10 & N@10 & M@10 \\
    \hline\hline

    \multirow{4}{*}{AOL4PS}
      & \multirow{2}{*}{Inactive}
      & \multirow{2}{*}{14.52}
      & \multirow{2}{*}{29.84}
      & ETEGRec-Q & 60.41 & 56.07 & 54.66 \\
      & & & & IntentGR & \textbf{69.74} & \textbf{65.05} & \textbf{63.52} \\
    \cline{2-8}

      & \multirow{2}{*}{Active}
      & \multirow{2}{*}{85.48}
      & \multirow{2}{*}{70.16}
      & ETEGRec-Q & 71.48 & 68.10 & 67.00 \\
      & & & & IntentGR & \textbf{79.39} & \textbf{75.91} & \textbf{74.77} \\
    \hline

    \multirow{4}{*}{AE}
      & \multirow{2}{*}{Inactive}
      & \multirow{2}{*}{39.51}
      & \multirow{2}{*}{39.88}
      & ETEGRec-Q & 21.36 & 15.14 & 13.18 \\
      & & & & IntentGR & \textbf{33.39} & \textbf{22.23} & \textbf{18.73} \\
    \cline{2-8}

      & \multirow{2}{*}{Active}
      & \multirow{2}{*}{60.49}
      & \multirow{2}{*}{60.12}
      & ETEGRec-Q & 23.08 & 16.36 & 14.26 \\
      & & & & IntentGR & \textbf{34.95} & \textbf{22.46} & \textbf{18.54} \\
    \hline
  \end{tabular}
  \end{adjustbox}
  \vspace{-5pt}
\end{table}

\subsubsection{Hyper-parameter Analysis}
\label{sec:hyper_param_analysis}

\begin{figure}[htbp]
  \centering

  \begin{minipage}[t]{0.48\columnwidth}
    \centering
    \begin{tikzpicture}
      \begin{axis}[
        width=\linewidth,
        height=0.72\linewidth,
        title={(a) Coefficient $\eta$},
        title style={font=\small\bfseries},
        xlabel={$\eta$ ($\times 10^{-4}$)},
        ylabel={Recall@10 (\%)},
        xmin=0.8, xmax=5.2,
        ymin=77.0, ymax=78.0,
        xtick={1,2,3,4,5},
        ytick={77.0,77.2,77.4,77.6,77.8,78.0},
        grid=major,
        grid style={draw=gray!55, line width=0.2pt},
        axis line style={black, line width=0.4pt},
        tick style={gray!55, line width=0.4pt},
        tick label style={font=\scriptsize},
        label style={font=\scriptsize},
        mark size=1.8pt,
      ]
        \addplot[
          color=orange!35,
          mark=*,
          mark options={fill=orange!35, draw=orange!35},
          line width=2.2pt
        ] coordinates {
          (1,77.48)
          (2,77.54)
          (3,77.46)
          (4,77.55)
          (5,77.40)
        };

        \addplot[
          color=orange,
          mark=*,
          mark options={fill=orange, draw=orange},
          line width=0.9pt
        ] coordinates {
          (1,77.48)
          (2,77.54)
          (3,77.46)
          (4,77.55)
          (5,77.40)
        };
      \end{axis}
    \end{tikzpicture}
  \end{minipage}
  \hfill
  \begin{minipage}[t]{0.48\columnwidth}
    \centering
    \begin{tikzpicture}
      \begin{axis}[
        width=\linewidth,
        height=0.72\linewidth,
        title={(b) Coefficient $\alpha$},
        title style={font=\small\bfseries},
        xlabel={$\alpha$},
        ylabel={Recall@10 (\%)},
        xmin=-0.05, xmax=1.05,
        ymin=77.0, ymax=78.2,
        xtick={0,0.2,0.4,0.6,0.8,1.0},
        xticklabels={$\approx0$,0.2,0.4,0.6,0.8,$\approx1$},
        ytick={77.0,77.2,77.4,77.6,77.8,78.0,78.2},
        grid=major,
        grid style={draw=gray!55, line width=0.2pt},
        axis line style={black, line width=0.4pt},
        tick style={gray!55, line width=0.4pt},
        tick label style={font=\scriptsize},
        label style={font=\scriptsize},
        mark size=1.8pt,
      ]
        \addplot[
          color=green!30,
          mark=*,
          mark options={fill=green!30, draw=green!30},
          line width=2.2pt
        ] coordinates {
          (0,77.75)
          (0.2,77.87)
          (0.4,77.82)
          (0.6,77.75)
          (0.8,77.99)
          (1.0,77.93)
        };

        \addplot[
          color=green!60!black,
          mark=*,
          mark options={fill=green!60!black, draw=green!60!black},
          line width=0.9pt
        ] coordinates {
          (0,77.75)
          (0.2,77.87)
          (0.4,77.82)
          (0.6,77.75)
          (0.8,77.99)
          (1.0,77.93)
        };
      \end{axis}
    \end{tikzpicture}
  \end{minipage}

  \caption{Hyper-parameter analysis on the coefficients of IntentGR.}
  \label{fig:hyper_params}
\end{figure}
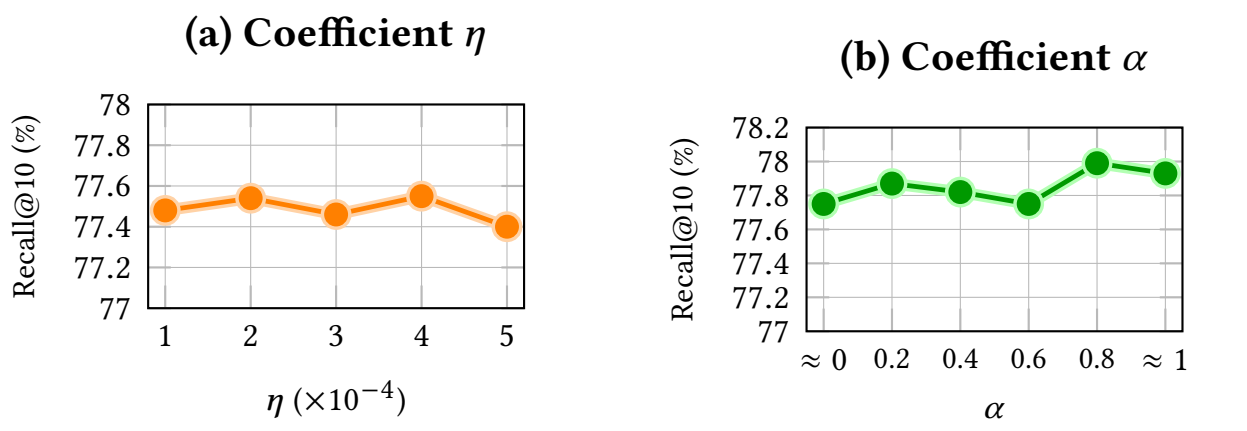

We study IntentGR's sensitivity to its two key hyper-parameters: the contrastive coefficient $\eta$ in QSA and the fusion coefficient $\alpha$ in QAM. 
1) \textbf{Coefficient in QSA.}
We evaluate IntentGR's performance under different values of the query-semantic contrastive loss coefficient $\eta$ in Eq.~\ref{loss_it_gr} and Eq.~\ref{loss_ft}. Specifically, we vary $\eta$ from $1\times10^{-4}$ to $5\times10^{-4}$ on the ETEGRec-Q+QBR+QSA setting and compare the corresponding Recall@10 on AOL4PS. As shown in Figure~\ref{fig:hyper_params}(a), IntentGR maintains stable performance across different $\eta$ values, with the highest Recall@10 obtained at $\eta=4\times10^{-4}$. This suggests that the proposed QSA is robust within a reasonable coefficient range. Therefore, we set $\eta=4\times10^{-4}$ in the final model. 
2) \textbf{Coefficient in QAM.}
We further analyze the fusion coefficient $\alpha$ in QAM under the full IntentGR setting. As defined in Eq.~\ref{eq:qam}, $\alpha$ controls the trade-off between the original non-query encoder states and the expanded query representation in the generation context. A larger $\alpha$ preserves more information from the original behavior context, while a smaller $\alpha$ strengthens the influence of query semantics during SID generation. We vary $\alpha$ within $\{10^{-6}, 0.2, 0.4, 0.6, 0.8, 1-10^{-6}\}$ and compare the corresponding Recall@10 on AOL4PS. As shown in Figure~\ref{fig:hyper_params}(b), $\alpha=0.8$ achieves the best Recall@10, suggesting that preserving sufficient behavior context while incorporating query semantics leads to the most effective fusion. The lower performance at boundary settings further indicates the necessity of balancing the two signals. Therefore, we set $\alpha=0.8$ in the final model.

\vspace{-0.5em}
\section{Conclusion}

We proposed IntentGR, an intent-enhanced end-to-end generative retrieval framework that strengthens query semantics across four complementary stages: input (QBR), optimization (QSA), training distribution (LQE), and decoding (QAM). QBR improves personalized search by introducing query-level behavioral evidence into the input, allowing the model to better capture the collective intent associated with a query. Once this enriched evidence is available, QSA, LQE, and QAM provide additional gains by aligning query semantics with the learned SID space, balancing the training distribution of long-tail queries, and strengthening query conditioning during decoding. Experiments on AOL4PS and the industrial AE dataset show that IntentGR consistently outperforms competitive baselines, with notable improvements on long-tail queries and inactive users. These results indicate that both enriched input evidence and effective integration mechanisms are important for building intent-aware end-to-end generative search systems. Future work may explore complementary evidence forms such as multimodal signal and session-level context to further enrich the intent signal.

\begin{acks}
This work was supported by the Fundamental and Interdisci-plinary Disciplines Breakthrough Plan of the Ministry of Education of China (JYB2025XDXM103).
\end{acks}

\section*{Ethics and Privacy Statement}

\textbf{Data Privacy.} This work uses two datasets: 1) AOL4PS is a publicly released academic benchmark for personalized search derived from the AOL query log. We use the public release and collect no additional information. Since query strings and timestamps are preserved verbatim, we treat the data as pseudonymized rather than fully anonymized: we make no attempt to re-identify individuals or link records to external sources, use the pseudonymous user identifiers only as offline grouping keys for aggregate statistics rather than as training input, and report only aggregate metrics without verbatim queries. We acknowledge the privacy concerns surrounding the original AOL release and use this benchmark solely as an offline research testbed.
2) AE is an industrial dataset collected from the e-commerce search platform AliExpress in accordance with its privacy policy and user agreement, and is stored and processed exclusively within the company's internal data warehouse under credentialed, project-scoped access. Each record consists of two sample-level identifiers, a pseudonymous user identifier (an opaque string that we cannot link to any real-world identity), the query text, the clicked item, and the user's recent interaction history represented as internal item IDs, with no directly identifying attributes such as contact details or payment information. None of the identifiers is used as model input---the model consumes only the query text and history item IDs mapped to SIDs. 
 
\textbf{Scope of the activity-level analysis.} User activity level is a \emph{post-hoc} analysis dimension only. On both datasets the model receives no user identifier and no activity signal at training or inference
time; the inactive/active partition is computed offline from interaction counts purely to slice the evaluation results.

\textbf{Generative Model.} Regarding model outputs, IntentGR generates SIDs within a fixed corpus rather than open-ended text, and thus poses no risk of producing harmful content. 

\textbf{Fairness.} By enriching supervision for long-tail queries, our method alleviates the head-query dominance of generative retrieval and improves service quality for low-frequency queries and inactive users, mitigating rather than amplifying popularity bias.

\bibliographystyle{ACM-Reference-Format}
\bibliography{sample-base}

\end{document}